\documentclass[british,pra,reprint,onecolumn,notitlepage]{revtex4-1}	% Note notitlepage to avoid page break
\usepackage[letterpaper]{geometry}
\usepackage{amssymb}
\usepackage{graphicx}
\usepackage[space]{grffile}
\usepackage{latexsym}
\usepackage{textcomp}
\usepackage{longtable}
\usepackage{multirow,booktabs}
\usepackage{amsfonts,amsmath,amssymb}
\usepackage{url}
\usepackage{hyperref}
\hypersetup{colorlinks=true,pdfborder={0 0 0},urlcolor=blue,citecolor=green,linkcolor=blue}
\usepackage[utf8]{inputenc}
\usepackage[english]{babel}



\begin{document}

\title{ePSproc: Post-processing suite for ePolyScat electron-molecule scattering calculations}

\author{Paul Hockett\\ National Research Council of Canada }

\maketitle

Software metapaper, structured for the \href{http://openresearchsoftware.metajnl.com}{Journal of Open Research Software (JORS)}.

Online version (Authorea) can be found at \href{https://www.authorea.com/users/71114/articles/122402/_show_article}{https://www.authorea.com/users/71114/articles/122402/\_show\_article}

This version: arXiv, 12th Nov. 2016.

% 12/11/16 - arxiv submission.
% 10/08/16 - This fork for review.

\section*{(1) Overview}

\section*{\vspace{0.5cm}}

\section*{ePSproc: Post-processing suite for ePolyScat electron-molecule scattering
calculations}

\section*{Paper Authors}

1. Hockett, Paul (\href{http://femtolab.ca}{femtolab.ca})

\section*{Paper Author Roles and Affiliations}

1. National Research Council of Canada, 100 Sussex Drive, Ottawa,
Ontario, Canada

\section*{Abstract}

ePSproc provides codes for post-processing results from ePolyScat
(ePS), a suite of codes for the calculation of quantum scattering
problems, developed and released by Luchesse \& co-workers \hyperref[csl:1]{(Gianturco et al. 1994)} \hyperref[csl:2]{(Natalense and Lucchese 1999)} \hyperref[csl:3]{(R. R. Lucchese and Gianturco 2016)}. ePS is a powerful
computational engine for solving scattering problems, but its inherent
complexity, combined with additional post-processing requirements,
ranging from simple visualizations to more complex processing involving
further calculations based on ePS outputs, present a significant barrier
to use for most researchers. ePSproc aims to lower this barrier by
providing a range of functions for reading, processing and plotting
outputs from ePS. Since ePS calculations are currently finding multiple
applications in AMO physics (see below), ePSproc is expected to have
significant reuse potential in the community, both as a basic tool-set
for researchers beginning to use ePS, and as a more advanced post-processing
suite for those already using ePS. ePSproc is currently written for
Matlab/Octave, and distributed via Github: \href{https://github.com/phockett/ePSproc}{https://github.com/phockett/ePSproc}.

\section*{Keywords}

\begin{enumerate}
\item Photoionization
\item Photoelectron angular distributions
\item Electron-molecule scattering
\item ePolyScat
\item AMO physics
\item Quantum physics
\item Dipole matrix elements
\item Matlab
\item Octave
\end{enumerate}

\section*{Introduction}

The \href{http://www.chem.tamu.edu/rgroup/lucchese/ePolyScat.E3.manual/manual.html}{ePolyScat (ePS) suite}, developed and released by Lucchese \& co-workers \hyperref[csl:1]{(Gianturco et al. 1994)} \hyperref[csl:2]{(Natalense and Lucchese 1999)} \hyperref[csl:3]{(R. R. Lucchese and Gianturco 2016)},
provides an open source software tool to the AMO physics community
for the calculation of a range of quantum scattering problems. At
heart, the codes use a Schwinger variational method to solve the continuum
electron wavefunction, for a given scattering potential and continuum
energy. ePS is written in Fortran 90, and makes use of MPI for parallel
processing. The code is modular, with an input file consisting of
a range of commands to define the specific calculation(s) of interest,
and typically produces a monolitic output file, consistent with the
traditional style of quantum chemistry software (e.g. Gamess, Gaussian
etc.). Some additional processing tools, and supplementary output
file options, are available to the user. User interaction is at the
command line or via text files. 

\begin{figure}[h!]
\begin{center}
\includegraphics[width=1\columnwidth]{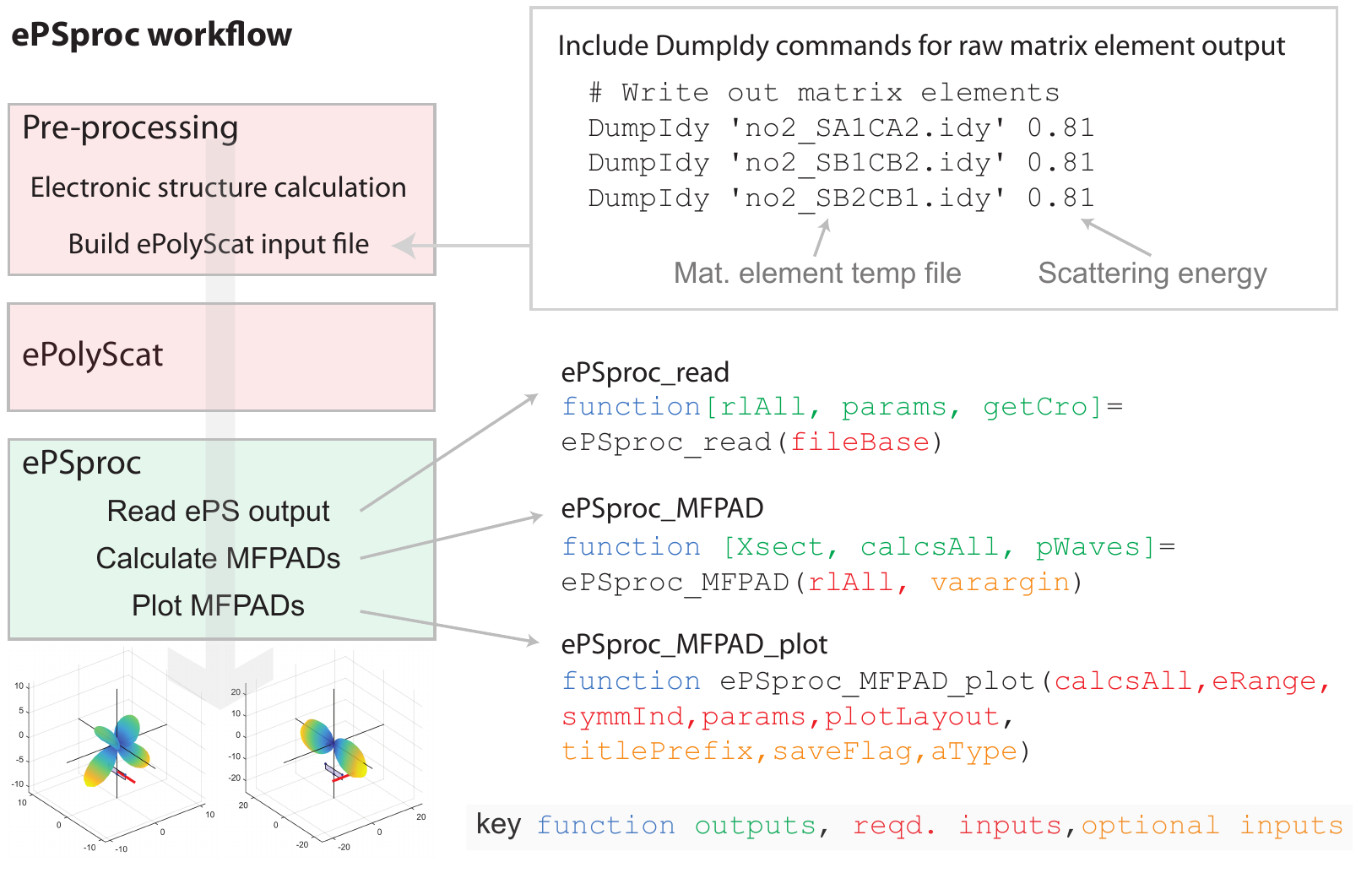}
\caption{{\label{fig:workflow}Overview of workflow for ePolyScat and ePSproc to calculate MFPADs. For export of raw matrix elements into ePS output file, the \textbf{DumpIdy} command is required in the input file for each set of records. The ePSproc function prototypes are shown in detail, further details of each function can be found at the heads of the \href{https://github.com/phockett/ePSproc}{source codes}.%
}}
\end{center}
\end{figure}

\href{https://github.com/phockett/ePSproc}{ePSproc, a suite of codes for post-processing of ePS outputs}, aims
to provide existing, or potential, ePS users with tools for:
\begin{itemize}
\item Parsing of ePS output files and extraction of segments/data of interest.
\item Visualization of ePS calculation results.
\item Additional post-processing, based on the raw matrix elements.
\end{itemize}
The workflow for ePolyScat and ePSproc is illustrated in figure \ref{fig:workflow}, and the full range of the current functionality of ePSproc is given in table \ref{tab:functions}, including some future aims. Full details of each function can be found in the \href{https://github.com/phockett/ePSproc}{source code and documentation}.

ePSproc evolved from the use of ePS for a variety of calculations,
in particular for molecular and laboratory frame photoelectron angular
distributions, and high-harmonic generation \hyperref[csl:4]{(Wörner et al. 2010)} \hyperref[csl:5]{(Bertrand et al. 2012)};
these works were, in turn, typically based on prior publications from ePS contributors developing
the underlying computational methods, e.g. \hyperref[csl:6]{(Toffoli et al. 2007)} \hyperref[csl:7]{(Le et al. 2009)}. In all cases,
ePS was utilized to calculate dipole matrix elements for a specific
problem, and significant post-processing was required to produce the
desired results. These various computations resulted in a range of
Matlab scripts and functions, for both basic and advanced processing
of ePS outputs. In this context ``basic'' functionality consists
of visualization of ePS results, and post-processing of the output
data in a manner similar to other tools within the ePS suite (e.g.
basic frame transformations), while ``advanced'' functionality consists
of making use of the raw matrix elements in further computations (e.g.
laboratory frame angular distributions for aligned molecular ensembles \hyperref[csl:8]{(Hockett 2015)}, Wigner time-delays \hyperref[csl:9]{(Hockett et al. 2016)} \hyperref[csl:10]{(Hockett 2016)}).
Aspects of these various existing, and carefully tested, codes have now been
consolidated into ePSproc. 

The first release of ePSproc (v1.0.0) is aimed at photoionization problems.
In this release, ePS output files are parsed, and the user can then
plot photoionization cross-sections and molecular frame photoelectron
angular distributions (MFPADs). The former are taken directly from the ePS
output, while the latter incorporates processing of the dipole matrix
elements output by ePS for calculation of the molecular frame photoelectron
scattering distributions for any arbitrary light polarization and molecular
frame alignment, as specified by the user.  The formalism employed is detailed in the following section.

Future releases of ePSproc will include calculations for a wider range
of cases, including time-dependent aligned distributions and high-harmonic generation. A port
of the software to the Python language, for a true open-source implementation,
is also planned.

\begin{table} 
\begin{tabular}{|c|c|p{10cm}|}
\hline 
Function & Implemented & Details\\
\hline 
\hline 
\verb|ePSproc_read|  & Y & Read information from main ePS output file. Reads details of molecular
geometry, scattering energies, cross-sections (\verb|GetCro|  segments) and
raw matrix elements (\verb|DumpIdy|  segments).\\
\hline 
\verb|ePSproc_MFPAD|  & Y & Calculate MFPADs for all input sets of matrix elements, for a given
polarization state and geometry \hyperref[csl:6]{(Toffoli et al. 2007)}.\\
\hline 
\verb|ePSproc_MFPAD_plot|  & Y & Plot results from ePSproc\_MFPAD.\\
\hline 
\verb|ePSproc_NO2_MFPADs_demo|  & Y & Basic script for running the example/benchmark calculations.\\
\hline 
\verb|ePSproc_LFPAD|  & N & Similar to ePSproc\_MFPAD, but for LFPADs.\\
\hline 
\verb|ePSproc_AFPAD|  & N & Calculate aligned frame (AF) PADs for a given molecular axis distribution \hyperref[csl:8]{(Hockett 2015)}.\\
\hline 
\verb|ePSproc_HHG|  & N & Calculate recombination matrix elements for use in HHG calculations \hyperref[csl:7]{(Le et al. 2009)}.\\
\hline 
\verb|ePSproc_wignerT|  & N & Calculate ionization (Wigner) time-delays \hyperref[csl:9]{(Hockett et al. 2016)}.\\
\hline 
\hline 
\end{tabular}
    \caption{{\label{tab:functions}List of main functions in ePSproc, implementation status in the \href{https://dx.doi.org/10.6084/m9.figshare.3545639.v1}{initial release v1.0.0}, and the purpose of the function. For full details, see the \href{https://github.com/phockett/ePSproc}{source code} and descriptions within. The references listed provide examples of each usage case, including the relevant theoretical formalism.}} 
\end{table}

\section*{Formalism}
Molecular frame photoelectron angular distributions (MFPADs) can be defined as per eqns. 1-3 of ref. \hyperref[csl:6]{(Toffoli et al. 2007)}:

\begin{equation}
I_{\mu_{0}}^{p_{i},p_{f}}(\theta_{\hat{k}},\phi_{\hat{k}},\theta_{\hat{n}},\phi_{\hat{n}})=\frac{4\pi^{2}E}{cg_{p_{i}}}\sum_{\mu_{i},\mu_{f}}|T_{\mu_{0}}^{p_{i}\mu_{i},p_{f}\mu_{f}}(\theta_{\hat{k}},\phi_{\hat{k}},\theta_{\hat{n}},\phi_{\hat{n}})|^{2}\label{eq:MFPAD}
\end{equation}

\begin{equation}
T_{\mu_{0}}^{p_{i}\mu_{i},p_{f}\mu_{f}}(\theta_{\hat{k}},\phi_{\hat{k}},\theta_{\hat{n}},\phi_{\hat{n}})=\sum_{l,m,\mu}I_{l,m,\mu}^{p_{i}\mu_{i},p_{f}\mu_{f}}(E)Y_{lm}^{*}(\theta_{\hat{k}},\phi_{\hat{k}})D_{\mu,-\mu_{0}}^{1}(R_{\hat{n}})\label{eq:TMF}
\end{equation}

\begin{equation}
I_{l,m,\mu}^{p_{i}\mu_{i},p_{f}\mu_{f}}(E)=\langle\Psi_{i}^{p_{i},\mu_{i}}|\hat{d_{\mu}}|\Psi_{f}^{p_{f},\mu_{f}}\varphi_{klm}^{(-)}\rangle\label{eq:I}
\end{equation}

In this formalism:
\begin{itemize}
\item $I_{l,m,\mu}^{p_{i}\mu_{i},p_{f}\mu_{f}}(E)$ is the radial part of
the dipole matrix element, determined from the initial ($i$) and final ($f$) state
electronic wavefunctions $\Psi_{i}^{p_{i},\mu_{i}}$and $\Psi_{f}^{p_{f},\mu_{f}}$,
photoelectron wavefunction $\varphi_{klm}^{(-)}$ and dipole operator
$\hat{d_{\mu}}$. The wavefunctions are indexed by irreducible
representation (i.e. symmetry) by the labels $p_{i}$ and $p_{f}$,
with components $\mu_{i}$ and $\mu_{f}$ respectively; $l,m$ are
angular momentum components, $\mu$ is the projection of the polarization
into the MF (from a value $\mu_{0}$ in the LF). Each choice of energy and irreducible
representation (and components thereof) corresponds to a single scattering calculation in ePolyScat, and produces a set of matrix elements $I^{(c)}_{l,m,\mu}$, where $(c)$ denotes that the results correspond to a single scattering calculation, and the matrix elements are again expanded in the spherical basis quantum numbers ${l,m,\mu}$. These are the raw matrix elements which ePSproc works with.
\item $T_{\mu_{0}}^{p_{i}\mu_{i},p_{f}\mu_{f}}(\theta_{\hat{k}},\phi_{\hat{k}},\theta_{\hat{n}},\phi_{\hat{n}})$
is the full matrix element (expanded in polar coordinates) in the
MF, where $\hat{k}$ denotes the direction of the photoelectron $\mathbf{k}$-vector,
and $\hat{n}$ the direction of the polarization vector $\mathbf{n}$
of the ionizing light. Note that the summation over components $\{l,m,\mu\}$
is coherent, and hence phase sensitive.
\item $Y_{lm}^{*}(\theta_{\hat{k}},\phi_{\hat{k}})$ is a spherical harmonic.
\item $D_{\mu,-\mu_{0}}^{1}(R_{\hat{n}})$ is a Wigner rotation matrix element,
with a set of Euler angles $R_{\hat{n}}=(\phi_{\hat{n}},\theta_{\hat{n}},\chi_{\hat{n}})$,
which rotates/projects the polarization into the MF .
\item $I_{\mu_{0}}(\theta_{\hat{k}},\phi_{\hat{k}},\theta_{\hat{n}},\phi_{\hat{n}})$
is the final (observable) MFPAD, for a polarization $\mu_{0}$ and
summed over all symmetry components of the initial and final states,
$\mu_{i}$ and $\mu_{f}$. Note that this sum can be expressed as
an incoherent summation, since these components are (by definition)
orthogonal.
\item $g_{p_{i}}$ is the degeneracy of the state $p_{i}$.
\end{itemize}

The laboratory frame photoelectron angular distribution (LFPAD) for a single molecular orientation can be similarly defined:

\begin{equation}
T_{\mu_{0}}^{p_{i}\mu_{i},p_{f}\mu_{f}}(\theta_{\hat{k}},\phi_{\hat{k}},\theta_{\hat{n}},\phi_{\hat{n}})=\sum_{l,m,\mu}I_{l,m,\mu}^{p_{i}\mu_{i},p_{f}\mu_{f}}(E)Y_{lm}^{*}(\theta_{\hat{k}},\phi_{\hat{k}})D_{\mu,\mu_{0}}^{1}(R_{\hat{n}})D_{m,m_{0}}^{l}(R_{\hat{n}})\label{eq:TLF}
\end{equation}

Which is similar to eqn. \ref{eq:TMF}, except with an additional
rotation matrix element to rotate the spherical harmonics into the
LF (defined by the polarization vector of the incident radiation), with components $m_{0}$.

As defined above, ePSproc works from the raw matrix elements, defined in sets corresponding to a single scattering calculation output from ePolyScat, $I^{(c)}_{l,m,\mu}$.  Application of the equations above, defined in terms of operations on each set of matrix elements and summation over sets, allows computation of the MF and LFPADs for any desired polarization geometry. Similar formalisms can be applied to obtain related quantities, such as recombination matrix elements for high harmonic generation, or Wigner ionization time-delays.

\begin{figure}[h!]
\begin{center}
\includegraphics[width=1\columnwidth]{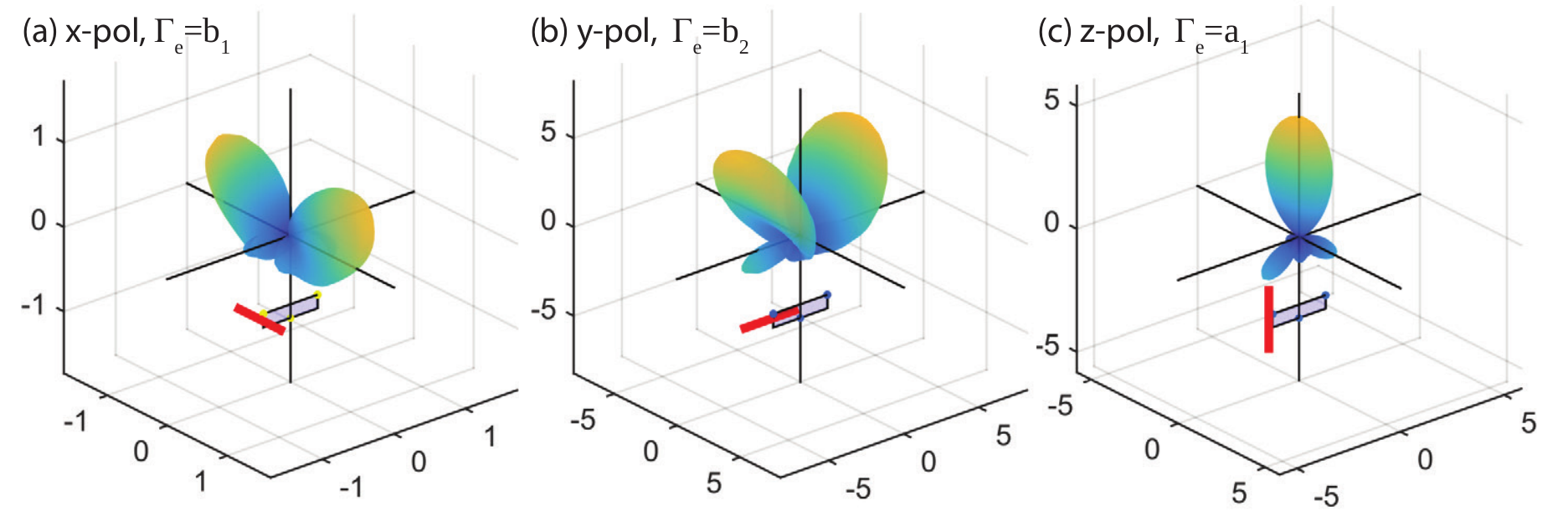}
\caption{{\label{fig:ePSprocDemo1}Example output from ePSproc. MFPADs for ionization of $NO_2$, via removal of a $(4a_1)$ electron at 22 eV (resulting in electron energies of 0.7eV), leading to the $(4a_1)^{-1}{}^3A_1$ state of $NO_2^+$ (c.f. figure 5 of ref. \hyperref[csl:6]{(Toffoli et al. 2007)}). Each panel shows the results for a single ePS calculation (set of matrix elements $I^{(c)}_{l,m,\mu}$) corresponding to a different polarization geometry. The polarization geometries are denoted as $x$, $y$ and $z$, corresponding to continuum symmetries $\Gamma_e=b_1,b_2,a_1$ respectively. Units are the angle-resolved cross-sections in MB.%
}}
\end{center}
\end{figure}

\begin{figure}[h!]
\begin{center}
\includegraphics[width=1\columnwidth]{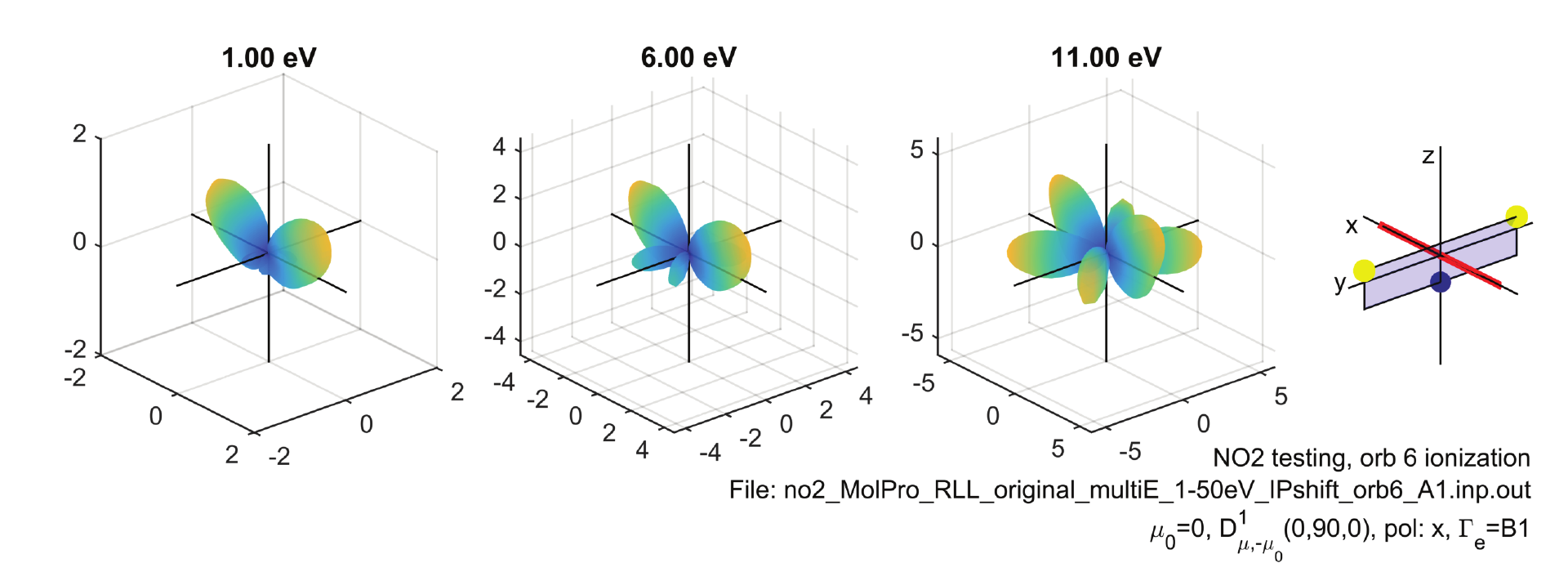}
\caption{{\label{fig:ePSprocDemo2}Example output from ePSproc. MFPADs for ionization of $NO_2$, via removal of a $(4a_1)$ electron, leading to the $(4a_1)^{-1}{}^3A_1$ state of $NO_2^+$ (as per figure \ref{fig:ePSprocDemo1}), for different continuum energies. The molecular structure and polarization geometry ($x$) is shown in the last panel. The text, as output by \textbf{ePSproc\_MFPAD\_plot()}, includes a user-supplied description \textbf{titlePrefix} (line 1), the source filename (line 2) and the (user-defined) calculation parameters for the E-field polarization in the LF $\mu_0$, frame rotation LF$\rightarrow$MF (defined by Euler angles $R_{\hat{n}}=(\phi_{\hat{n}},\theta_{\hat{n}},\chi_{\hat{n}})$) and continuum symmetry $\Gamma_e$. Units are the angle-resolved cross-sections in MB.%
}}
\end{center}
\end{figure}

\section*{Quality control}

\begin{figure}[h!]
\begin{center}
\includegraphics[width=1\columnwidth]{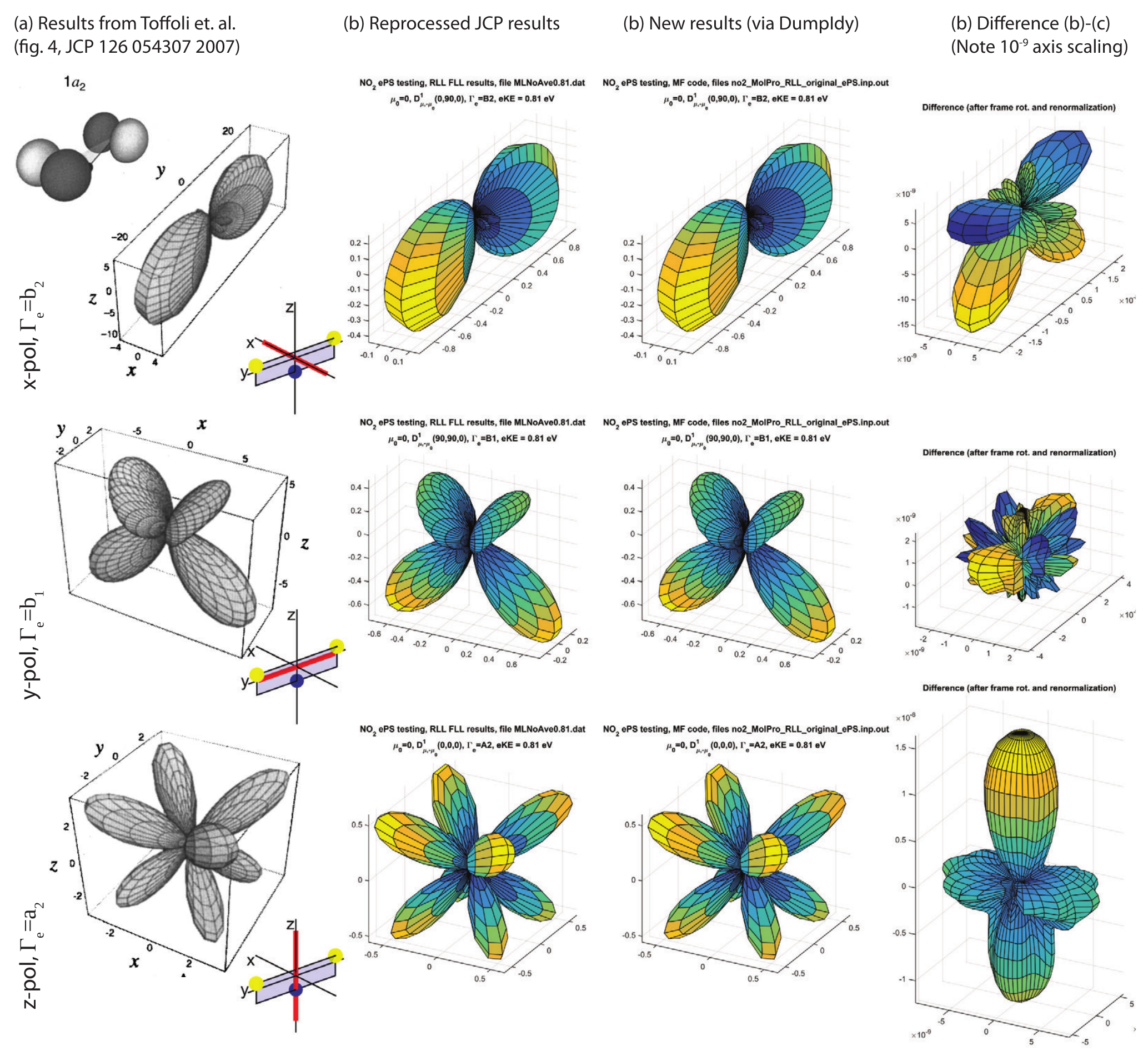}
\caption{{\label{fig:ePSproc-testing}ePSproc benchmarking. Comparison between (a) previously published ePS results (reproduced from fig. 4 of ref. \hyperref[csl:6]{(Toffoli et al. 2007)}), for ionization of $NO_2$ at 14.4 eV, and ejection of a $(1a_2)$ electron - the ionizing orbital is shown top left; (b) reprocessed MFPADs from the original data, output by ePolyScat's OrientCro function (original results provided by R.R. Lucchese); (c) comparison results using ePSproc; (d) difference between results (b) and (c), note the $10^{-9}$ scaling, indicating excellent agreement. Each row shows the results for a different polarization geometry, $x$, $y$ and $z$, corresponding to continuum symmetries $\Gamma_e=b_2,b_1,a_2$ respectively. See main text for further details.%
}}
\end{center}
\end{figure}

As well as the existing scientific publications listed above, which
indicate the lineage of the current suite, the release version of
ePSproc has been carefully tested for the generation of MFPADs against benchmark calculations provided by R.R. Lucchese
(one of the original ePS authors, as described above), based on calculations originally published in ref. \hyperref[csl:6]{(Toffoli et al. 2007)}. An overview of the testing and results is provided in figure \ref{fig:ePSproc-testing}.

For the testing procedure, input matrix elements for ePSproc were
obtained from an ePS calculation (version E3), running under Ubuntu
LTS 14. The electronic structure inputs for this calculation were
the sample $NO_{2}$ Gaussian output, provided with the ePS.E3 release \hyperref[csl:3]{(R. R. Lucchese and Gianturco 2016)},
although a new ePS input file (including matrix element output via \verb|DumpIdy|  commands) was created for this computation (see figure \ref{fig:workflow} for visualization of this workflow). Benchmark
results, provided by R.R. Lucchese, originated from the output of
the ePS \verb|OrientCro|  routine (which provides a data-file of angle-resolved
cross-sections, expanded as harmonics, for a given polarization state),
and were post-processed using Matlab code also provided by R.R. Lucchese.
The resulting $(\theta,\phi)$ gridded data was then plotted (figure \ref{fig:ePSproc-testing}(b)). The reprocessing was verified against the previously published results, figure \ref{fig:ePSproc-testing}(a). 

The new calculations, with post-processing in ePSproc, are shown in figure \ref{fig:ePSproc-testing}(c), and were directly compared to the (reprocessed) benchmark results via numerical subtraction; the results of this comparison are shown in figure \ref{fig:ePSproc-testing}(d), indicating
agreement with the benchmark calculations to $10^{-9}$. In this case, the
comparison is very good, although it is of note that this testing
does not differentiate between differences in the raw ePS outputs
(which could arise for many reasons, both computational and due to
differences in the input files) and differences in the post-processing.

The ePSproc
results shown were obtained on an 64-bit Intel Core I7 system, running
Matlab R2010b under Ubuntu LTS release 14, and similar verification
has been successfully obtained using a 64-bit Intel Core I5 system,
running Matlab R2015a under Windows 7. (The results have not been verified
with Octave at the current time.)

The current release of ePSproc includes the required ePS output file
and demo script to reproduce the benchmark post-processing, and should
be the first test run by a new user (users running ePS can, additionally, test the ePS portion of the benchmark calculations, via changes to the ePS input file). The demo calculation also provides
the opportunity for the user to further test ePSproc post-processing and visualization
by, for instance, testing alternative polarization geometries. Additional
details are provided in the documentation accompanying ePSproc.

\section*{(2) Availability}

\vspace{0.5cm}

\section*{Operating system}

Any OS capable of running Matlab or Octave. File parsing currently
makes use of Windows or Linux command-line tools for searches, so may require small
modification for other operating systems.

\section*{Programming language}

The code suite is written for Matlab/Octave. No Matlab proprietary
or toolbox functions are used in ePSProc, so any version of Matlab
or Octave should be suitable. The current version has been tested
in Matlab releases R2010b, R2013a, R2015a.

\section*{Additional system requirements}

None.

\section*{Dependencies}

None - no additional libraries required.

\section*{List of contributors}

\begin{itemize}
\item P. Hockett - development of ePSProc.
\item R. R. Lucchese, N. Sanna, A. P. P. Natalense, and F. A. Gianturco  - development and release of ePS \hyperref[csl:1]{(Gianturco et al. 1994)} \hyperref[csl:2]{(Natalense and Lucchese 1999)} \hyperref[csl:3]{(R. R. Lucchese and Gianturco 2016)}.
\end{itemize}
Special thanks to R.R. Lucchese, for general help with ePS usage and
providing benchmark ePS outputs and post-processed MFPADs for verification
of ePSproc outputs.

\section*{Software location:}
{\bf Archive}
\begin{description}	
\item[Name:] Figshare 	
\item[Persistent identifier:] \href{https://dx.doi.org/10.6084/m9.figshare.3545639.v1}{DOI: 10.6084/m9.figshare.3545639}
\item[Licence:] CC-BY 	
\item[Publisher:]  P. Hockett 	
\item[Version published:] Initial release, v1.0.0 	
\item[Date published:] Release 17/04/16, archived 07/08/16
\end{description}
{\bf Code repository} 
\begin{description}
\item[Name:] GitHub 	
\item[Persistent identifier:] \href{https://github.com/phockett/ePSproc}{https://github.com/phockett/ePSproc}
\item[Licence:] GNU GPL	
\item[Date published:] 17th April 2016 (v1.0)
\end{description}

\section*{Language}
English

\section*{(3) Reuse potential}
As detailed above, the use of ePS has recently become widespread, with users in a range of related areas of AMO physics. It is anticipated that many of these users would find a use for ePSproc. Potential new users will, hopefully, find that the bar to entry for ePS is lowered since, at a basic level, ePSproc provides an easy, and well-tested, methodology for bringing raw matrix elements from ePS into Matlab. This provides a rapid and transparent means of analyzing and visualizing ePS results, and removes the requirement for researchers to develop and test their own post-processing techniques. Beyond this, there are a range of potential usage cases, from the basic exploration of MFPADs in order to build physical intuition, to more detailed and extensive computations considering, e.g. molecular dynamics, aligned ensembles, HHG, etc. (see table \ref{tab:functions} for some examples). In release 1.0.0 of ePSproc only MFPAD calculations are implemented, and additional usage cases will be incorporated in future releases. The applications listed in table \ref{tab:functions} have already been developed, but still require code rationalization and integration into ePSproc before release. It is anticipated that users will also develop post-processing routines for other usage cases, making use of the low-level ePS I/O and computational functions implemented in ePSproc, and these additions could be incorporated into future releases to further extend the capabilities of the code.

Additional future plans to extend ePSproc include the creation of a library of sample calculations (initially based on the publications cited above), and porting the code to a fully open-source framework, probably the python language.

User support for ePSproc can be obtained directly from the author, via the \href{https://github.com/phockett/ePSproc}{project GitHub page} or by email.

\section*{Acknowledgements}
PH thanks the multiple collaborators and co-authors - many of whom are listed in the references - who encouraged and suggested the cavilier use of ePS ``out of the box", for many different problems incorporating electron scattering and photoionization. This spirit of ``shoot first, ask questions later" indeed raised many questions which, eventually, necessitated proper use of ePS and careful post-processing of the results, and sharpened related foundational expertise - efforts well worth making.

\section*{Competing interests}
The authors declare that they have no competing interests.

\section*{References}
\phantomsection
\label{csl:1}Gianturco FA, Lucchese RR, Sanna N (1994) \href{http://dx.doi.org/10.1063/1.467237}{Calculation of low-energy elastic cross sections for electron-{CF}4 scattering}. The Journal of Chemical Physics 100:6464. doi: 10.1063/1.467237

\phantomsection
\label{csl:2}Natalense APP, Lucchese RR (1999)  \href{http://dx.doi.org/10.1063/1.479794}{Cross section and asymmetry parameter calculation for sulfur 1s photoionization of SF6}. The Journal of Chemical Physics 111:5344. doi: 10.1063/1.479794

\phantomsection
\label{csl:3}R. R. Lucchese APPN N. Sanna, Gianturco FA (2016) Lucchese group \& ePolyScat website, 

\href{http://www.chem.tamu.edu/rgroup/lucchese/}{http://www.chem.tamu.edu/rgroup/lucchese/}. 

\phantomsection
\label{csl:4}Wörner HJ, Bertrand JB, Hockett P, and others (2010) \href{http://dx.doi.org/10.1103/physrevlett.104.233904}{Controlling the Interference of Multiple Molecular Orbitals in High-Harmonic Generation}. Phys Rev Lett. doi: 10.1103/physrevlett.104.233904

\phantomsection
\label{csl:5}Bertrand JB, Wörner HJ, Hockett P, and others (2012) \href{http://dx.doi.org/10.1103/physrevlett.109.143001}{Revealing the Cooper minimum of N2 by Molecular Frame High-Harmonic Spectroscopy}. Phys Rev Lett. doi: 10.1103/physrevlett.109.143001

\phantomsection
\label{csl:6}Toffoli D, Lucchese RR, Lebech M, and others (2007) \href{http://dx.doi.org/10.1063/1.2432124}{Molecular frame and recoil frame photoelectron angular distributions from dissociative photoionization of NO2}. The Journal of Chemical Physics 126:054307. doi: 10.1063/1.2432124

\phantomsection
\label{csl:7}Le A-T, Lucchese RR, Tonzani S, and others (2009) \href{http://dx.doi.org/10.1103/physreva.80.013401}{Quantitative rescattering theory for high-order harmonic generation from molecules}. Phys Rev A. doi: 10.1103/physreva.80.013401

\phantomsection
\label{csl:8}Hockett P (2015) \href{http://dx.doi.org/10.1088/1367-2630/17/2/023069}{General phenomenology of ionization from aligned molecular ensembles}. New Journal of Physics 17:023069. doi: 10.1088/1367-2630/17/2/023069

\phantomsection
\label{csl:9}Hockett P, Frumker E, Villeneuve DM, Corkum PB (2016) \href{http://dx.doi.org/10.1088/0953-4075/49/9/095602}{Time delay in molecular photoionization}. J Phys B: At Mol Opt Phys 49:095602. doi: 10.1088/0953-4075/49/9/095602

\phantomsection
\label{csl:10}Hockett P (2016) \href{https://dx.doi.org/10.6084/m9.figshare.3545639.v1}{Figshare: Time Delay in Molecular Photoionization (inc. Supp. Mat. \& data)}. 

\end{document}